\documentstyle[aps,epsfig,twocolumn]{revtex}
\begin{document}
\draft
\flushbottom
\twocolumn[  
\hsize\textwidth\columnwidth\hsize\csname @twocolumnfalse\endcsname


\author{Oleg Zaitsev$^1$, R. Narevich$^2$, and R. E. Prange$^1$}
\address{$^1$Department of Physics, University of Maryland, College Park, Maryland\\
20742\\
$^2$Department of Physics and Astronomy, University of Kentucky, Lexington,\\
Kentucky 40506}
\title{Quasiclassical Born-Oppenheimer approximations}
\maketitle

\tightenlines
\widetext
\advance\leftskip by 57pt
\advance\rightskip by 57pt  

\begin{abstract}
We discuss several problems in quasiclassical physics for which approximate
solutions were recently obtained by a new method, and which can also be
solved by novel versions of the Born-Oppenheimer approximation. These cases
include the so-called bouncing ball modes, low angular momentum states in
perturbed circular billiards, resonant states in perturbed rectangular
billiards, and whispering gallery modes. Some rare, special eigenstates,
concentrated close to the edge or along a diagonal of a nearly rectangular
billiard are found. This kind of state has apparently previously escaped
notice.
\end{abstract}

\vspace{7mm}
]
\narrowtext
\tightenlines

\section{Introduction}

A major success of the quasiclassical method began with Martin Gutzwiller. With his
famous `trace formula'\cite {Gutz} he started on the road to quantize `hard' chaotic
systems, ones possessing only strongly unstable orbits. The work he did and the work
he inspired can now be said to have solved the quantum spectral problem for two
dimensional hard chaos systems.

The study of Gutzwiller's problem led to the introduction of new tools such
as the dynamical zeta function\cite{zeta}. The surface of section transfer
operator (SSTO), a generalization of the boundary integral method, is an
important technique invented by Bogomolny\cite{bogolss}. This operator, 
$T(E),$ is a Fredholm kernel which carries a ray of energy $E$
quasiclassically from one intersection with a surface of section to the
next. It was first employed as a powerful reformulation of the Gutzwiller
trace formula. The spectrum is given by the zeroes of the Fredholm
determinant $D(E)=\det \left[ 1-T(E) \right]$ which is an expression of the
dynamical zeta function\cite{zeta}. The trace formula itself is given by the
logarithmic derivative of $D(E),$ expanded in traces of powers of $T.$ These
traces are expressed quasiclassically in terms of periodic orbits.

However, chaos is not required to use the SSTO. It was discovered\cite
{raysplit},\cite{pnz} that sometimes the equation 
\begin{equation}
T(E)\psi =\psi  \label{Teq0}
\end{equation}
could be solved quasiclassically for both the energy [equivalent to $D(E)=0$%
] and the `surface of section' [SS] wavefunction $\psi .$ The
two-dimensional wavefunction is obtained from $\psi $. In this way we found
wavefunctions and energy levels for some problems that had been extensively
studied by numerics and trace formulas\cite{circle},\cite{flux},\cite{Smil},%
\cite{KapHel}, but which had not been suspected of having simple analytic
solutions. After obtaining these solutions, we realized that they could
often also be found to the same level of accuracy by extensions of the
Born-Oppenheimer approximation (BOA)\cite{BO}. The BOA is not usually
regarded as quasiclassical, of course. Further research revealed that these
extensions of the BOA are related to the `parabolic equation' and etalon
methods mentioned below. This paper is devoted to illustrating these
relations.

The problems susceptible to this method are ones which are `locally' `nearly
integrable'. By this we mean that there is, for these problems, a region of
phase space where the {\em short }orbits are close to those of some
integrable system. This often gives rise to rare {\em special states}, with
striking properties. Perhaps the best known example is the `whispering
gallery' idea of Lord Rayleigh\cite{LR} which explains quantum phenomena\cite
{quantwg} as well as effects in acoustics and seismics, optics\cite{optics}
and radio propagation.

A number of methods have been developed to deal with such states, where it
is as important to understand the wavefunctions as it is the energies. Among
these methods are the Keller-Rubinow or ray method\cite{Keller}, the
parabolic equation method (PEM) of Leontovich\cite{Leon} and Fock\cite{Fock}%
, and the etalon method of Babich and Buldyrev\cite{BB}. However, there is
no method so systematic that it is sure that all such special states have
already been found.

The Keller-Rubinow method is purely quasiclassical and is based on first
understanding the classical mechanics, or `rays' in the optics nomenclature.
It sorts out the local integrability, exploiting caustics or an adiabatic
invariant of the motion. The last two methods mentioned start from the basic
partial differential equation [e.g. Schr\"odinger or Helmholtz] defining the
problem, which is rescaled in appropriate variables and approximated.
Systematic corrections to the leading order are also studied, and indeed
seem to be the focus of much of the theory. These latter methods are related
to the BOA, a fact not previously noted.

\section{Born-Oppenheimer Approximation}

The Born-Oppenheimer or adiabatic approximation is fundamental to the
quantum theory of molecules as well as to the quantum theory of the solid
state. It treats the electrons' position{\bf \ }${\bf r}_e$ as `fast'
variables , compared to the `slow' ionic positions ${\bf R}_i$. This is
based on the small electron-ion mass ratio $m_e/M_i.$

Most texts give a simple formulation of the BOA \cite{Baym}. Schr\"odinger's
equation is, in terms of electronic and ionic positions ${\bf r}_e,$ ${\bf R}%
_i$, 
\begin{eqnarray}
&&\left[ -\frac{\hbar ^2}{2m_e}\nabla _e^2+V({\bf r}_e,{\bf R}_i)-\frac{%
\hbar ^2}{2M_i}\nabla _i^2\right] \Psi ({\bf r}_e{\bf ,R}_i)  \nonumber
\label{S} \\
&=&E\Psi ({\bf r}_e{\bf ,R}_i).  \label{S}
\end{eqnarray}
The Born-Oppenheimer Ansatz is 
\begin{equation}
\Psi ({\bf r}_e{\bf ,R}_i)=\Phi ({\bf r}_e|{\bf R}_i)\psi ({\bf R}_i).
\label{BO}
\end{equation}
where $\Phi $ is, say, the $N$'th eigenstate in the electronic variables
which solves 
\begin{equation}
\left[ -\frac{\hbar ^2}{2m_e}\nabla _e^2+V({\bf r}_e,{\bf R}_i)\right] \Phi (%
{\bf r}_e|{\bf R}_i)=U({\bf R}_i)\Phi ({\bf r}_e|{\bf R}_i).  \label{SPhi}
\end{equation}
In $\Phi ,$ ${\bf R}_i$ is treated as a parameter, i.e., the quantum motion
of the fast variable ${\bf r}_e$ is found for a fixed value of the slow
variable $\ {\bf R}_i$. The `potential' $U$ is an energy eigenvalue of this
`fast' part of the Schr\"odinger equation, often but not necessarily that of
the electron ground state. The adiabatic invariance is invoked by the
assumption that the electrons' quantum state label $N$ does not change as $%
{\bf R}_i$ is slowly varied.

Now use Eq. (\ref{BO}) in Eq. (\ref{S}), neglect derivatives of $\Phi $ with
respect to ${\bf R}_i,$ multiply by $\bar \Phi ({\bf r}_e|{\bf R}_i)$, and
integrate over ${\bf r}_e$. This gives for the slow variables ${\bf R}_i$
the equation 
\begin{equation}
\left[ -\frac{\hbar ^2}{2M_i}\nabla _i^2+U({\bf R}_i)\right] \psi ({\bf R}%
_i)=E\psi ({\bf R}_i),  \label{Spsi}
\end{equation}
which completes the leading order solution.

\section{Imperfect square cavity}

As a first almost trivial example with apparently new results, consider a
two-dimensional microwave or laser or quantum-dot cavity which is not quite
square. For example, with Dirichlet's conditions, take a trapezoid with
boundaries $x=0,$ $x=L,$ $y=L,$ $y=\epsilon x$ where $\epsilon $ is small.
Consider states whose wavelength $\lambda $ is short, i.e. $\lambda <<L.$
Even if the perturbation is so small as to satisfy $\epsilon L<<\lambda ,$
there can be states which differ drastically from the states of a perfect
square. The condition for such novelty is $\sqrt{\epsilon }L\geq \lambda .$

The trapezoid problem was studied formally in Morse and Feshbach\cite{MF},
and is also the basis of recent work on quantum chaos\cite{KapHel}, but no
special states are mentioned. Look for a state fast in the $y$ direction,
slow in $x.$ Take $L=1$, and choose for the Ansatz, $\Phi (y|x)\propto \sin
\left[ n\pi \left( 1-y\right) /\left( 1-\epsilon x\right) \right] $ with $%
n\pi $ large, satisfying $\partial ^2\Phi /\partial y^2=$ `constant' $\times
\Phi $ and vanishing at $y=1$ and $y=\epsilon x$. The `constant' is of
course a relatively slowly varying function of $x.$ Then $U(x)=\left[ n\pi
/\left( 1-\epsilon x\right) \right] ^2\approx n^2\pi ^2+2\epsilon n^2\pi
^2x. $ The resulting equation for $\psi $ is familiar, 
\begin{equation}
-\psi ^{\prime \prime }+\alpha ^3x\psi =E_m\psi  \label{PsiAiry}
\end{equation}
where $E_m=E-n^2\pi ^2$ and $\alpha =(2\epsilon n^2\pi ^2)^{1/3}.$ The
solution is (if $\alpha >>1$) an Airy function, $\psi (x)=%
\mathop{\rm Ai}
(\alpha x-z_m)$, where $E_m=\alpha ^2z_m.$ Let $%
\mathop{\rm Ai}
(-z_m)=0,$ to make $\psi (0)=0.$ If $\alpha -z_m>>1,$ $\psi $ also
effectively vanishes at $x=1.$

Thus we have a number of rather simple special wavefunctions, labelled by $%
n,m$, which are concentrated along the long edge of the trapezoid. Namely, 
\begin{equation}
\Psi _{n,m}\approx \sin \left[ n\pi \left( 1-y\right) /\left( 1-\epsilon
x\right) \right] 
\mathop{\rm Ai}
(\alpha x-z_m)  \label{Psitrap}
\end{equation}
We show in Figs. 1, 2 a couple of representations of these states, for $%
n=55, $ $m=1,2.$ We choose $\epsilon =0.01,$ rather less than $\lambda =.036,
$ which in turn is rather less than $\sqrt{\epsilon }=0.1.$ The wave
functions were obtained numerically and are compared with Eq. (\ref{Psitrap}%
). Presumably a result like this, so fundamental to the construction of
cavities, is in an engineering textbook somewhere, but if so, we haven't
found it.

\vspace{2mm}
\begin{figure}[tbp]
{\hspace*{0.2cm}\psfig{figure=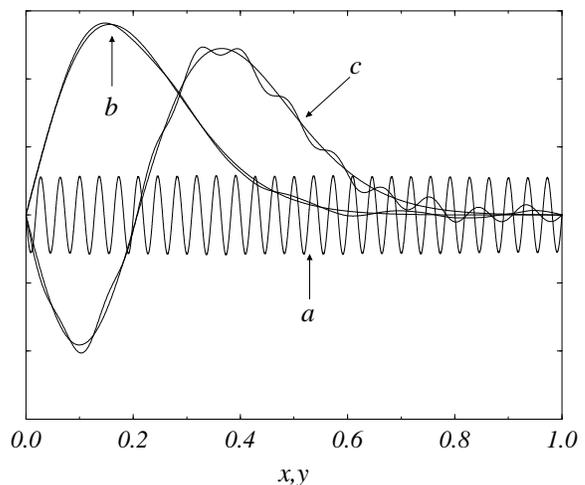,height=7.5cm,width=6.1cm,angle=270}}
{\vspace*{.16in}}
\caption[ty]{Numerically obtained states for the nearly square trapezoidal
billiard, $\epsilon =.01$, $n=55$, $m=1,2.$ a) $\Psi _{55,1}(x=0.01,y)$ vs $%
y $; b) $\Psi _{55,1}(x,y=.99)$ and $%
\mathop{\rm Ai}
(\alpha x-z_1)$ vs $x$; c) $\Psi _{55,2}(x,y=.99)$ and $%
\mathop{\rm Ai}
(\alpha x-z_2)$ vs $x.$ The magnitude of the states has been normalized.}
\end{figure}

It is not necessary to have an explicit small parameter. Keller\cite{Keller}
obtained approximate energies for `bouncing ball' states in convex
billiards. Even more dramatic results were obtained by the BOA some time ago%
\cite{Taylor} explaining the `bouncing ball' modes observed numerically\cite
{MK} in the Bunimovich stadium billiard. These results may be readily be
generalized to a distorted stadium. Following Primack and Smilansky\cite
{Smil}, take a `slanted' stadium billiard in which the radius of the `left'
semicircle is $R(1-\epsilon /2),$ and the right radius is $R(1+\epsilon /2).$
The two endcaps are separated by a distance $2a.$ The BOA is $\Phi
(y|x)=\sin \left[ n\pi \frac{y+R-\xi (x)}{2(R-\xi (x))}\right] $. Here $%
y=R-\xi (x)$ is the upper boundary of the billiard. This gives a potential
for the slow equation $V(x)=\left( n^2\pi ^2/2R^3\right) \xi (x).$ For the
standard stadium, $\xi =0$ for $\left| x\right| <a$ and becomes positive
outside that region. If $n^2\pi ^2$ is large, $V(x)$ rises rapidly for $%
\left| x\right| >a,$ so it is a sort of square well potential. For the
slanted stadium, the bottom of the well $V(x)$ is sloped, i.e. $\xi (x)=%
\frac 12\epsilon xR/a$ for $\left| x\right| <a,$ and again the potential
rises rapidly outside that region. The effect of this slope depends on $a/R$
and the slow quantum number $m$ as well as $n.$ Thus, if the change of
`potential energy' $V(a)-V(-a)\propto $ $\left( n^2/R^2\right) \epsilon
>>\left( m/a^2\right) ,$ the spacing of the slow energy levels, there will
be a large effect on the slow wavefunctions, concentrating them in the wider
part of the billiard, similar to the trapezoidal billiard above. With the
opposite inequality, the slanting of the sides of the billiard can be
neglected.

\vspace{2mm}
\begin{figure}[tbp]
{\hspace*{0.2cm}\psfig{figure=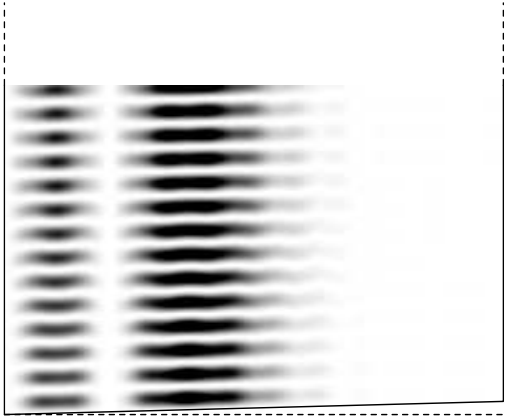,height=6.8cm,width=7.5cm,angle=0}}
{\vspace*{.13in}}
\caption[ty]{Density plot of $\left| \Psi _{55,2}(x,y)\right| ^2,$ the same state
as in Fig. 1c. The state is shown in the lower quarter of the billiard, $%
y<0.25,$ and the scale is expanded in the $y$ direction by a factor 3 in
order to show more details of the wave function. A dashed line is shown at $%
y=0.$ }
\end{figure}

This smooth transition of the quantum levels
contrasts with the mathematics
of the classical periodic orbits used in the trace formula. With parallel
sides to the billiard, there is a set of nonisolated periodic orbits
contributing a very large term to the trace formula. An arbitrarily small
slope mathematically eliminates all such nonisolated periodic orbits.
However, for a bounce or two, they remain close to the ideal periodic orbits
of the channel with two parallel sides.

Reference\cite{Smil} sorts this out with respect to the trace formula. It is
remarkable that the eigen{\em states }and {\em energies }start to be
substantially modified when $\epsilon \geq \left( R/an\pi \right) ^2$ while
the low terms of the trace formula and thus the so-called {\em length
spectrum,} are only modified if $\epsilon \geq R/an\pi .$ In other words,
the {\em low terms }of the trace formula are essentially unmodified if the
actual billiard is a ideal stadium billiard with `optically flat' errors
much smaller than a wavelength. This is however, not good enough to
guarantee that the wavefunctions are close to the ideal wavefunctions.

\section{Bogomolny operator}

The preceding results can be obtained by a method\cite{pnz} based on
Bogomolny's\cite{bogolss} quasiclassical surface of section transfer
operator, $T$. The main idea is to organize the phases which appear in the
transfer operator and its eigenfunctions according to their rate of change
as a function of position on the surface of section. Thus, there are `fast'
and `slow' parts to the Bogomolny equation.

In the trapezoid problem above, we take the space part of the surface of
section to be $y=1.$ The SSTO for billiards is 
\begin{equation}
T(x,x^{\prime }|k)=\sqrt{\frac 1{2\pi i\hbar }\left| \frac{\partial
^2S(x,x^{\prime })}{\partial x\partial x^{\prime }}\right| }e^{i\frac{%
S(x,x^{\prime })}\hbar }.  \label{Tdef}
\end{equation}
Here the action, $S(x,x^{\prime })=\hbar kL(x,x^{\prime })$, is that of the
classical orbit leaving the surface of section at $x^{\prime }$ and
returning to it for the first time at $x.$ It is expressed by $L(x,x^{\prime
}),$ the length of this path. The wavenumber $k$ $=$ $\sqrt{E}$ in our
units. We focus on orbits in the neighborhood of the $(1,0)$ resonance
orbits of the unperturbed square. Such an orbit starts at ${\bf r}^{\prime }%
{\bf =(}x^{\prime },1)$ and returns to ${\bf r}=(x,1)$ after making one
bounce from the bottom at approximately ${\bf \bar r=}\left( \frac 12%
(x+x^{\prime }),\frac 12\epsilon (x+x^{\prime })\right) $. For small $%
x-x^{\prime },$ $L\approx 2+\frac 14(x-x^{\prime })^2-\epsilon (x+x^{\prime
}).$

The second term of $kL$ is rapidly varying with $x^{\prime }$, the last term
is relatively slow. We solve Eq. (\ref{Teq0}), i. e. $\int dx^{\prime
}T(x,x^{\prime }|k)\psi (x^{\prime })=\psi (x),$ by making the Ansatz $\psi
(x^{\prime })=\exp \left[ \pm ik\sqrt{\epsilon }F(x^{\prime })\right] $.
This function, which has an intermediate rate of variation, followed by
stationary phase approximation for the integral, causes the important values
of $x^{\prime }$ to be close to $x,$ and allows $F(x^{\prime })\approx
F(x)+(x^{\prime }-x)f(x).$ Solution of Eq. (\ref{Teq0}) requires $k\sqrt{%
\epsilon }F=\frac 23\left[ z_m-\alpha x\right] ^{3/2},$ where $\alpha $ is
defined in the previous section. This gives for $\psi $ the WKB
approximation to the Airy function.

In short, the simplest version of this SSTO technique gives quasiclassically
the same $\psi $ as the BOA. Knowing $\psi ,$ the full wave function is
given by $\Psi ({\bf r)=}\int dxG_0\left[ {\bf r,r(}x)\right] \psi (x),$
where the kernel $G_0$ is related to the free space Green's function between
a point ${\bf r}$ in the interior of the billiard and the point ${\bf r(}x)$
on the surface of section. The result is essentially the BOA.

Note that the SSTO gives first the surface of section wavefunction, then the
BOA wavefunction, reversing the order of the Born-Oppenheimer approximation.
The BOA is simpler and more intuitive, when it works.

\section{Diagonal states}

We next give a result for which we cannot construct a simple BOA Ansatz.
Consider the same trapezoidal billiard but now look for special states
related to the $(1,1)$ periodic orbits of the square. Then, it's not too
hard to work out the SSTO by making the Ansatz $\psi (x^{\prime })=\exp (\pm
ikx^{\prime }/\sqrt{2})\exp \left[ \pm ik\sqrt{\epsilon }F(x^{\prime
})\right] $. The first phase factor chooses orbits at approximately 45$%
^{\circ }$ to the square sides. In this case, the result is that there are
states, induced by the small perturbation, that are concentrated along the
diagonal $(0,0)\rightarrow (1,1)$ but {\em not }along $(0,1)\rightarrow
(1,0) $. [In an appropriate parameter range, we do expect states with
`scars' along that diagonal, however.] Such a state, obtained numerically,
is shown in Figs. 3 and 4. We have found the 2-dimensional state $\Psi (x,y)$
to order $\epsilon ,$ but it is too cumbersome to display here. The
following wave function (for even $n$) vanishes on the boundary, and is the
same as our theoretical wave function to order $\epsilon ^0$ (but not $%
\epsilon $). It is not an adequate BOA Ansatz but it is simple and it gives
the gross features of the result. The function is $\Psi (x,y)\approx \cos
\left[ \frac 12n\pi (x+\tilde y)\right] u(x-\tilde y)-\cos \left[ \frac 12%
n\pi (x-\tilde y)\right] u(x+\tilde y)$ where $u(x)=%
\mathop{\rm Ai}
(\alpha \left| x\right| -z_0)$ for $\left| x\right| \leq 1$, extended with
period two outside this range, and $\tilde y=(y-\epsilon x)/(1-\epsilon x)$.
Here $z_0\approx 1.02$ is the first maximum of the Airy function. The
energies of such states are given by $E_{n,m}=\frac 12\left( n\pi \right) ^2+%
\frac 12\alpha ^2z_m$ where $z_m$ is a zero or extremum of the Airy
function. We display the remarkable nodal structure of such a state in Fig.
5.

\vspace{2mm}
\begin{figure}[tbp]
{\hspace*{0.2cm}\psfig{figure=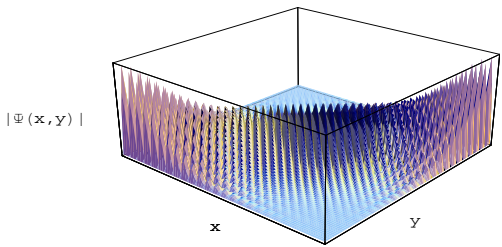,height=4.2cm,width=7.5cm,angle=0}}
{\vspace*{.13in}}
\caption[ty]{A three-dimensional representation of a state of the trapezoid
concentrated near the long diagonal. The theoretical state is plotted. $%
\epsilon =0.03,$ $n=50,$ $m=0.$}
\end{figure}

With a little practice, it is easy to foresee when special states exist or
not. For example, there are no states concentrated along the diagonal for
the symmetric trapezoid billiard with sides $x=0,1$, $y=\epsilon x,$ $%
1-\epsilon x$, but there are states concentrated along the side $x=0$. For
the parallelogram billiard with sides $x=0,1$, $y=\epsilon x$, $y=1+\epsilon
x$, there are states along the long diagonal but no states near the edges.

\section{Extended zone scheme}

The BOA requires a choice of fast and slow variables. In the preceding
example, a reflection of the wave incident on a boundary at 45$^{\circ }$
switches the roles of $x$ and $y.$ The following device overcomes this
difficulty, although it is cumbersome to apply to the example of the
preceding section.

Consider a perturbed rectangular billiard. The perturbation can be a
potential, a magnetic field, or magnetic flux lines. We assume the
perturbation is classically weak, meaning that the classical orbit does not
change much because of the perturbation in one traversal of the system. Our
SSTO technique automatically gives the result\cite{npzmag}, where pictures
of some special states may be found.

For the BOA, we choose the important example where a magnetic field perturbs
the $(1,1)$ resonance of the rectangle. First, we consider an auxiliary
problem, a method of images, from whose solutions the desired answer is
constructed.

The auxiliary problem extends the domain of the wavefunction to all of two
dimensional space. The perturbing magnetic flux, originally defined only in
the rectangle, is reflected about each of the sides of the rectangle, and
then the process is repeated so that the result forms a lattice of period $%
2a $ in the $x$-direction, and period $2b$ in the $y$-direction. The vector
potential is chosen to have this periodic structure.

\vspace{2mm}
\begin{figure}[tbp]
{\hspace*{0.2cm}\psfig{figure=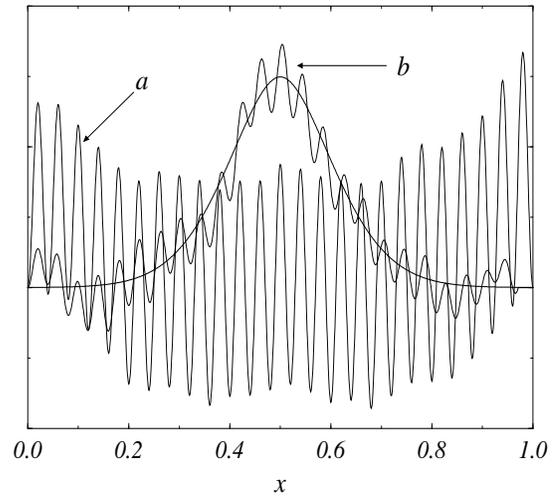,height=7.5cm,width=6.9cm,angle=270}}
{\vspace*{.13in}}
\caption[ty]{The same state as shown in Fig. 3. The numerically obtained state is
plotted, basically along the diagonals of the trapezoid. a) $\Psi (x,x)$, b) 
$\Psi (x,1-x).$ Also shown in part b) is the theoretical wavefunction, which
is smoother. Part b) is plotted for $0<x<1/(1+\epsilon ).$}
\end{figure}

Next we find the high energy wavefunctions of a periodic lattice, where the
lattice potential is classically weak. The methods developed in the study of
channelling\cite{channelling} are appropriate and are effectively the BOA.
We start with the $(1,1)$ channel, which amounts to assuming that the fast
direction is given by the variable $\xi =(ax+by)/c$ and the slow by $\eta
=(bx-ay)/c$, where $c=\sqrt{a^2+b^2}$. Take charge $e=1$, mass = $\frac 12$,
and $\hbar =1$ and rewrite Schr\"odinger's equation in variables $\xi $, $%
\eta $ as 
\begin{eqnarray}
&&\left\{ \left[ -i\partial _\xi -A_\xi (\xi ,\eta )\right] ^2+\left[
-i\partial _\eta -A_\eta (\xi ,\eta )\right] ^2\right\} \Psi (\xi ,\eta ) 
\nonumber  \label{xieta} \\
&=&E\Psi (\xi ,\eta ).  \label{xieta}
\end{eqnarray}
The channelling approximation averages along the fast, $\xi ,$ direction.
This amounts to assuming $\Phi (\xi |\eta )=\exp (ik\xi )$ and replacing $%
A_\xi (\xi ,\eta )$ by $\bar A_\xi (\eta )=\int_0^{L_\xi }d\xi A_\xi (\xi
,\eta )/L_\xi .$ Here $L_\xi =2c$. Because $k=-i\partial _\xi $ is large,
the terms $A_{\xi ,}^2$ $A_\eta ^2$ can be neglected in comparison to $k\bar 
A_\xi (\eta )\ $and $\bar A_\eta $ vanishes. The remaining $\eta $
dependence is obviously periodic.

\vspace{2mm}
\begin{figure}[tbp]
{\hspace*{2.0cm}\psfig{figure=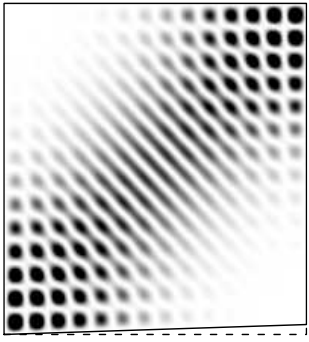,height=5.0cm,width=5.0cm,angle=0}}
{\vspace*{.13in}}
\caption[ty]{A density plot of a state similar to that shown in Figs. 3,4, with $%
n=28.$ The dashed line is $y=0.$}
\end{figure}

Take the origin at the rectangle center, for symmetry reasons. Let $\psi
(\eta )=u(c\eta /b-a/2)=u(x-ay/b-a/2)$. Then $u$ satisfies 
\begin{equation}
-u_m^{\prime \prime }(x)+V(x)u_m(x)=E_mu_m(x)  \label{um}
\end{equation}
where 
\begin{equation}
V(x)=-2%
{\textstyle {b^2 \over c^2}}
k\bar A_\xi \left( 
{\textstyle {bx \over c}}
+%
{\textstyle {ab \over 2c}}
\right) .  \label{V_A}
\end{equation}
The reason for the shift in the definition of $u$ is that then $V(x)=V(-x)$,
and $u$ will be simple on the boundaries where the Dirichlet conditions are
imposed.

The integral defining $\bar A_\xi $ is the flux enclosed by the periodic
orbit in the original rectangle, and so is independent of gauge. For uniform
field $B$, $\bar A_\xi (\eta )=+B\eta (1-c\eta /ab)$ for $0\leq \eta <ab/c$, 
$\bar A_\xi (\eta )=-B(2ab/c-\eta )(c\eta /ab-1)$ for $ab/c\leq \eta <2ab/c$%
, $\bar A_\xi (\eta +2ab/c)=\bar A_\xi (\eta ).$ In terms of $x$, $%
V(x)=-(Bkb^3a/c^3)\left[ \frac 12-2(x/a)^2\right] $ for $\left| x\right| <%
\frac 12a$,$\,$ repeated antiperiodically outside this region. The solution
of Eq. (\ref{um}) is a Bloch state, which satisfies 
\begin{equation}
u_m(x+2a)=e^{i\beta }u_m(x),  \label{Bloch}
\end{equation}
where $m$ labels the `band' and $\beta $, the `crystal momentum'. The band
index $m$ is a quantum number and $\beta $ is yet to be determined. The
energy is given by 
\begin{equation}
E=k^2+c^2E_m/b^2.  \label{E_m}
\end{equation}

Quantum states will be strongly affected by the flux if the dimensionless
parameter $Bk(ab/c)^3$ is large. This parameter is $2\pi (\phi /\phi
_0)(kc)\left( ab/c^2\right) ^2$, where $\phi $ is the flux $\phi =abB$, and $%
\phi _0=2\pi \hbar /e$ is the flux quantum.

We therefore find a set of approximate solutions to the plane problem,
labeled by $k,\,m$ and $\beta $, $\Psi _I(x,y)=\exp \left[
ik(ax+by)/c\right] u_m(x-ay/b-a/2)$. Given $\Psi _I,$ another solution is
given by the symmetry under rotation by $\pi $, $\Psi _{III}(x,y)=\Psi
_I(-x,-y)$. Two other solutions are given by $\Psi _{II}(x,y)=\exp \left[
ik(-ax+by)/c\right] u_m(-x-ay/b+a/2)$ and $\Psi _{IV}(x,y)=\Psi _{II}(-x,-y)$%
. All four of these channels clearly have the same energy.

The solution for the original rectangle can be constructed from these four
solutions, provided some quantization conditions are met. We consider two
subspaces, $r=\pm 1$, even or odd under $(x,y)\rightarrow (-x,-y).$ The
state can be written in the form 
\begin{equation}
\Psi _{nm}=A\left( \Psi _I+r\Psi _{III}\right) +B(\Psi _{II}+r\Psi _{IV}).
\label{Psinm1}
\end{equation}
Imposing the condition $\Psi _{nm}\left( \pm \frac 12a,y\right) =0$ yields $%
A=-B\exp (-i\chi _a)$, and $\exp (i\beta )=\exp (-2i\chi _a)$ where $\chi
_a=ka^2/c.$ The condition $\Psi _{nm}\left( x,\pm \frac 12b\right) =0$ gives 
$A=-rB\exp (i\chi _b)$ and $\exp (i\beta )=\exp (2i\chi _b)$ where $\chi
_b=kb^2/c.$ Therefore, $\chi _a+\chi _b=\pi n$, where $n$ is even for $r=1,$
and odd for $r=-1.$ This yields 
\begin{equation}
k=\frac{\pi n}c.  \label{kn}
\end{equation}
One finds $\beta =2\chi _b-2\pi n_b=-2\chi _a+2\pi n_a.$ We may assume that $%
\left| \beta \right| \leq \pi ,$ which fixes the integers $n_{a,b}$ as the
integer part of $\left( \chi _{a,b}\pi ^{-1}+1/2\right) $. Then $%
n_a+n_b=ck/\pi =n.$

For a square, $a=b$, there is a further symmetry, and $\beta $ is either $0$
or $\pi ,$ that is, the appropriate state is either at the band top or band
bottom. For an arbitrary ratio $a/b$, $k\xi =k(ax+by)/c$ is not of the form $%
\pi n_ax/a+\pi n_by/b.$ However, the $n_{a,b}$ just found are the integers
which most closely satisfy this relation.

Another case of great interest is that of an Aharonov-Bohm flux line in an
integrable billiard\cite{ABLine}. We show in Fig. 6 current streamlines of a
state for the case of a unit square, $a=b=1,$ with a flux line at its
center. The potential is this case is $V(x)=-C,$ $\left| x\right| <\frac 12,$
$V(x)=+C,$ $\left| x-1\right| <\frac 12,$ repeated with period $2.$ Here $C=$
$\left( 2\pi \phi /\phi _0\right) k/\sqrt{8}$. In Fig. 6 we have taken $\phi
/\phi _0=$ $0.1$, the flux in units of the flux quantum, to be small in
order to minimize diffraction effects. The state is not much localized
spatially, but the current is very regular. For symmetry reasons, the
structure is particularly simple along a coordinate axis, as we display in
the figure. Because of diffraction, our theory of the flux line case has
relatively large corrections. Nevertheless, our simple theory captures the
main features of many states. We have related results\cite{NPZPhysaE}
published elsewhere.

In this case of a perturbed rectangular billiard, we have been able to find
a bigger problem which admits a decomposition into fast and slow variables
and thus a Born-Oppenheimer Ansatz. The original problem is solved by a
superposition of results of the auxiliary problem.

\section{ Fast and slow asymptotics}

We now give an example in which a separation of variables into fast and slow
does not hold over the whole domain, but does work over a limited domain,
that is, however, sufficient to solve the problem.

Consider a weakly distorted unit circle billiard. This is defined in polar
coordinates by the boundary $r=1+\epsilon \Delta R(\theta )$, where $%
\epsilon $ is small, $\Delta R\,$ is of order unity, and $\Delta R$ does not
vary too rapidly. A number of papers have appeared on this topic recently%
\cite{circle}.

\vspace{2mm}
\begin{figure}[tbp]
{\hspace*{0.2cm}\psfig{figure=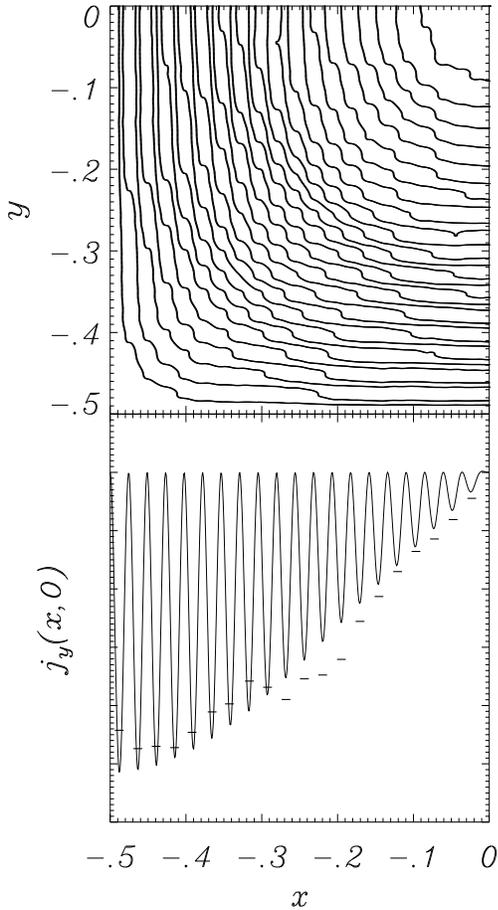,height=12.9cm,width=7.5cm,angle=0}}
{\vspace*{.13in}}
\caption[ty]{Upper figure: Numerical persistent current streamlines of a state
induced by an Aharonov-Bohm flux line containing 0.1 flux quanta, located at
the center of a square billiard. A quarter of the billiard is shown. Lower
figure: Current in the $y$-direction on the $y$ axis. The theoretical
current is plotted. The numerical current is similar but has minima at the
depths shown by the horizontal lines. The quantum numbers are $n=82,$ $m=0.$}
\end{figure}

We confine attention to high energy states whose classical counterparts pass
close to the center of the billiard. Such states have low angular momentum,
that is, dimensionless angular momenta $\nu <<k$. This suggests that we
consider the radial variable $r\,$ to be `fast', and the angular variable $%
\theta $ to be `slow' and thus the Ansatz $\Phi =J_{\nu (\theta )}(kr)$.
This, however, does not work. The reason is that near the origin, $r$ is not
fast compared with $\theta .$

However, we need to know $\Phi $ only near the boundary in order to impose
the Dirichlet conditions. Without a condition at the origin, using the
asymptotic expansion for Bessel's functions suggests the Ansatz 
\begin{equation}
\Phi (r|\theta )\approx \frac 1{\sqrt{kr}}\cos \left[ kr+\frac{\nu (\theta
)^2-\frac 14}{2kr}+\alpha (\theta )\right]  \label{Phiasym}
\end{equation}
where $\nu (\theta )$ is the order of the Bessel functions and $\alpha
(\theta )$ is a phase which mixes the asymptotic Bessel and Neumann
functions. The slow equation is $\psi ^{\prime \prime }+\nu ^2\psi =0.$ To
make $\Phi $ vanish at $r=1+\epsilon \Delta R(\theta )$ requires 
\begin{equation}
k+k\epsilon \Delta R(\theta )+\frac{\nu (\theta )^2-\frac 14}{2k}+\alpha
(\theta )=(n-\frac 12)\pi .  \label{argPhi}
\end{equation}

It is necessary to determine two functions, $\nu $ and $\alpha $, from Eq. (%
\ref{argPhi}). We expect $\nu $ to depend on $\frac 12\left[ \Delta R(\theta
)+\Delta R(\theta \pm \pi )\right] \equiv \overline{\Delta R(\theta )}$, and 
{\em not} on $\delta R(\theta )\equiv \frac 12\left[ \Delta R(\theta
)-\Delta R(\theta \pm \pi )\right] .$ Classically, a low angular momentum
state sees equally both sides of the circle, if at $\theta ,$ then also at $%
\theta \pm \pi .$ To achieve this, we take $\alpha (\theta )=-k\epsilon
\delta R(\theta )+\alpha _0,$ where $\alpha _0$ is to be determined. This
implies that 
\begin{equation}
\nu (\theta )^2=E_m-V(\theta )  \label{VCirc}
\end{equation}
where $V(\theta )=2k^2\epsilon \overline{\Delta R(\theta )}.$ This acts as a
potential in the slow equation, and it will be important if $k\sqrt{\epsilon 
}>>1.$

Again there is a Schr\"odinger equation with a periodic potential $V(\theta
)=V(\theta +\pi ),$ and thus solutions labelled by the band index $m$ and
`crystal momentum' $\beta .$ Since physically $\psi (\theta +2\pi )=\psi
(\theta )$,\thinspace $\beta =0$ or $\pi .$ The constant $\alpha _0$ is
determined from the special case $\Delta R=0,$ which implies $E_m=m^2,$ with 
$m$ integer, $\beta =\pi \left( m%
\mathop{\rm mod}
2\right) .$ For $\Delta R=0,$ the solution $\Phi (r|\theta )=J_m(kr)$
implies $\alpha _0=\frac 12\beta -\pi /4.$

In Fig. 7 we show a state for a stadium billiard as in Sec. III, with a
short straight side of length $2a=2\epsilon =0.02$ and end radius $R=1.$
Then $\Delta R=\epsilon \left| \cos \theta \right| +O(\epsilon ^2).$ Some
other states are shown in Ref.\cite{pnz}.

\vspace{2mm}
\begin{figure}[tbp]
{\hspace*{0.2cm}\psfig{figure=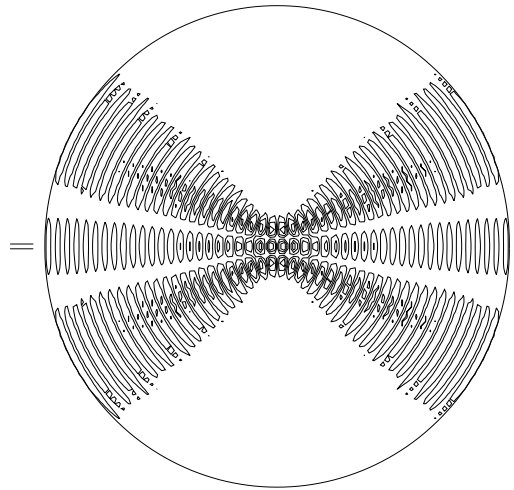,height=7.3cm,width=7.5cm,angle=0}}
{\vspace*{.13in}}
\caption[ty]{Numerically obtained contour plot of a state of a Bunimovitch
stadium billiard with end cap radius unity.whose straight sides are short,
of length $2a,$ where $a=0.01.$ The parallel lines near the left side of the
boundary have separation $2a.$ This state has $m=3,$ i.e. two transverse
nodes, and $n=25.$}
\end{figure}

This application of the BOA differs from others in that we do not have a
separation into slow and fast variables over the whole system, but we do
have such a separation, asymptotically, over the part of the system that
matters most.

\section{Whispering gallery modes}

Quantization of higher order periodic orbit resonances in nearly circular
billiards cannot be done easily by a Born-Oppenheimer method. However, the
whispering gallery limit can be handled with a modification of the BOA, and
is not restricted to the nearly circular case. Rather, we assume only a two
dimensional smooth convex billiard.

The results have long been known, but the methods we introduce are easier
and more intuitive than earlier techniques. These modes correspond to
classical motion which stays close to the billiard boundary while rapidly
moving along the boundary. The effect was discussed by Lord Rayleigh, first
for rays in his {\em Theory of Sound, }and later in terms of waves\cite{LR}.
Lord Rayleigh did not consider the eigenmodes, however, which were first
quantized by Keller\cite{Keller} as an example of what came to be known as
EBK quantization. Keller's work was based on the assumption of the existence
of caustics in the corresponding classical motion. This was proved later by
Lazutkin\cite{Lazutkin} who obtained an adiabatic invariant.

We first obtain the result by SSTO. The surface of section is the billiard
boundary. Let $s$ be the distance along the boundary. The ray goes from $%
s^{\prime }$ to $s,$ with $s^{*}=s-s^{\prime }$ small.

Let the radius of curvature at $\bar s=\frac 12(s+s^{\prime })$ be $R(\bar s%
).$ The curvature varies slowly in the sense $dR/d\bar s$ never gets too
large and $R$ is always positive and finite. Then, we use Eq. (\ref{Tdef})
above and approximate 
\begin{equation}
L(s,s^{\prime })\simeq 2R(\bar s)\sin \frac 12\left| \frac{s-s^{\prime }}{R(%
\bar s)}\right| \approx \left| s-s^{\prime }\right| -\frac{\left|
s-s^{\prime }\right| ^3}{24R(\bar s)^2}.  \label{LW}
\end{equation}
We now look for a solution of Eq. (\ref{Teq0}). $\,$ We take as Ansatz 
\begin{equation}
\psi (s)=\exp \left\{ ik\left[ s-F(s)\right] \right\}  \label{psiW}
\end{equation}
where $dF/ds\equiv f(s)<<1.$ Thus, apart from the explicit $\exp (iks)$, $%
\psi (s)$ varies relatively slowly.

We do the integral $T\psi $ by stationary phase. [The BOA below improves on
this.] The stationary phase point is $s^{\prime }=s-s^{*}$ where $\left(
s^{*}\right) ^2/8R^2=f$. It can be checked later that the $s^{\prime }$
dependence of $R$ can be neglected. Doing the integral gives 
\begin{equation}
T\psi =i\exp \left[ i\frac 23kR(2f)^{3/2}\right] \psi (s).  \label{Tpsi}
\end{equation}
A solution requires 
\begin{equation}
f=f_m(s)=F_m^{\prime }=\left[ \frac{3\pi }{\sqrt{8}kR(s)}(m+\frac 34)\right]
^{\frac 23}.  \label{fm}
\end{equation}
This is the first quantization condition. The condition that $f$ be small
requires \thinspace $m<<kR(s).$

The other quantization condition is a result of the requirement that $\psi $
is single valued on the boundary, that is, $\psi (0)=\psi ({\cal L})$ where $%
{\cal L}$ is the circumference of the billiard. This condition can be
expressed 
\begin{equation}
k\left[ {\cal L}-F_m({\cal L})+F_m(0)\right] =2\pi n  \label{Fn}
\end{equation}
Thus the quantization depends on $\int_0^{{\cal L}}R(s)^{-2/3}ds,$
effectively the mean 2/3 power of the curvature.

We now turn to the BOA. The billiard is locally a circle of radius $R(s).$
If $R$ is constant, the solution is $\Psi =J_\nu (kr)e^{i\nu \theta }$ where 
$J_\nu $ is a Bessel function. We assume $\nu $ is large. [This is close to
the treatment of Lord Rayleigh, who, however, takes for granted that the
index $\nu $ is an integer. Most of the rest of his discussion concerns the
asymptotic properties of Bessel functions of large index, a subject still
under development at the time. The etalon method also considers this Bessel
function, but does not distinguish fast and slow variables.]

We assume that for a more general convex billiard $\Psi $ has this form
locally. Namely we make the Ansatz 
\begin{equation}
\Psi ({\bf r})=\Phi (\rho |s)\psi (s)=J_{\nu (s)}(kr_s)\psi (s)
\label{BesselA}
\end{equation}
where $r_s$ is a radial coordinate from the local center of curvature. Let
the variable $\rho \geq 0$ be given by $r_s=R(s)-\rho .$ With the exception
of the factor $\exp (iks)$ all $s$ dependence is slow compared with the
`fast' variable $\rho .$ The index $\nu $ can vary smoothly and there is no
reason to make it an integer. We could already guess $\psi (s)=\exp \left[
i\int^sd\bar s\,\nu (\bar s)/R(\bar s)\right] $ where we have replaced the
local angle variable $d\theta _s$ by $ds/R.$ Parametrize $\nu =kR(1-f)$
where $f$ is small (and turns out to be the same $f$ introduced earlier).
Using a standard asymptotic formula\cite{AS} we have

\begin{eqnarray}
J_{\nu (s)}\left( k\left[ R(s)-\rho \right] \right) &=&J_\nu \left[ \nu +\nu
^{\frac 13}\left( \frac{kRf-k\rho }{\nu ^{\frac 13}}\right) \right] 
\nonumber \\
\ &\simeq &\left( \frac 2{kR}\right) ^{\frac 13}%
\mathop{\rm Ai}
\left[ -2^{\frac 13}\frac{kRf-k\rho }{\left( kR\right) ^{\frac 13}}\right] .
\label{BesselExp}
\end{eqnarray}
We can replace $\nu ^{1/3}$ by $(kR)^{1/3}$ when multiplied with the small
quantities $f$ or $\rho .$

The first quantization condition is that $\nu (s)$ must be chosen to solve $%
J_{\nu (s)}\left[ kR(s)\right] =0.$ A convenient analytic approximation is
obtained from Eq. (\ref{BesselExp}), again in terms of the zeroes $z_m$ of
the Airy function, 
\begin{equation}
f=z_m/2^{\frac 13}(kR)^{\frac 23}.  \label{fAiry}
\end{equation}

This improves on Keller's treatment or the $T$ operator above which is
equivalent to approximating $%
\mathop{\rm Ai}
(-x)\propto \sin (\frac 23x^{\frac 32}+\frac \pi 4),$ valid for large $x.$

To find $\psi (s)$ we substitute the Ansatz of Eq. (\ref{BesselA}) into the
Helmholtz equation, neglecting derivatives of $J_\nu $ with respect to $s.$
The angular derivative term $(1/r_s)^2\partial ^2/\partial \theta _s^2$ can
be replaced by $\partial ^2/\partial s^2$ so we see that $\psi ^{\prime
\prime }(s)+(\nu ^2/R^2)\psi =0.$ This has the approximate solution given in
Eq. (\ref{psiW}) above. From this, the requirement that the wavefunction be
single valued on the boundary gives the second quantization condition. Thus
the solution is 
\begin{equation}
\Psi (s,\rho )\approx C(s)%
\mathop{\rm Ai}
\left( -z_m+\frac{k\rho }{\left[ kR(s)\right] ^{\frac 13}}\right) \psi (s)
\label{PsiAiry}
\end{equation}
The slowly varying prefactor $C(s)$ is a normalization of the Airy function
determined by current conservation.

The validity of the BOA requires that the argument of the Airy function vary
more rapidly with $\rho $ than with $s$ [in the region that the Airy
function is large]. This condition can be written $\left| R^{\prime }\right|
<<\left( kR\right) ^{2/3}/z_m.$ The existence of a classical constant
replacing $k^{2/3}/z_m$ in this strong inequality is the condition for the
existence of a caustic. The caustic is at the zero argument of the Airy
function, i.e. at $\rho =z_m(kR)^{1/3}/k.$ Thus if $R$ is finite and smooth,
there is always a sufficiently large energy such that states remaining close
to the boundary exist.

If $1/R$ vanishes at some point, it is known that caustics do not exist, and
the BOA must fail near that point. However, numerical evidence (at
relatively low energies, of course) suggests that whispering gallery states
exist even when the curvature vanishes. No detailed theory has been advanced
for this case, to our knowledge. We have made some progress on this problem,
which will be published elsewhere.

\section{Conclusions}

With slight extensions, the Born-Oppenheimer approximation can, rather
quickly and easily, give the leading order results for a number of
interesting states. These include the important and well-known cases of the
whispering gallery modes, and the bouncing ball states. It also includes
some states only recently uncovered by the SSTO method. It is relatively
easy to apply when, at least asymptotically, the variables can be separated
into fast and slow. This can happen because of a small parameter, or just
because the particular state has that special property.

The PEM and etalon methods are similar to the BOA in the sense that they
deal directly with the partial differential equations. The detailed
procedures and motivations for each step is rather different from the BOA.
The leading results are the same, however. An advantage of the BOA is that
it is simpler and taught in standard courses on quantum mechanics while the
other methods are less well known.

These methods are not {\em a priori }semiclassical. However, if a variable
is fast because it has more energy than the slow variable, semiclassical
methods can be used. The Keller-Rubinow method is classic, but a little
cumbersome in practice. The SSTO method seems to be in a certain sense more
general than the Born-Oppenheimer. Compared with Keller-Rubinow, it has the
advantage of the simplifications coming from use of the surface of section.
Compared with BOA, it first finds the slow wavefunction, and from that the
fast one. Often the slow wavefunction is more interesting physically.

Another well known semiclassical method sometimes compared with the SSTO
method is the Birkhoff-Gustavson normal form\cite{BG}. However,
Birkhoff-Gustavson is not adapted to finding special wavefunctions. It
rather gives a large number of wavefunctions and energy levels at once, and
is similar to a numerical method in that respect.

Formal corrections to the PEM and etalon methods have been written down\cite
{BB}, which differ from the usual methods used with the BOA. These
corrections are not much used in practice. Rather, one resorts to a
numerical method. Moreover, quite large corrections to the wavefunctions are
sometimes found in the numerics. This is because the theory singles out some
relatively small subset of special states which are approximately decoupled
from all the rest. Nothing prevents some unrelated state from having nearly
the same energy as the special state. The residual coupling then mixes the
two appreciably. The energies are given quite well, although not necessarily
as accurately as the mean level spacing of all the levels.

Although it is sometimes possible to find approximations to both states and
to estimate the mixing parameters, it is hard to do that systematically. It
is perhaps more advisable to think of the wavefunctions obtained as being
those of {\em quasimodes}\cite{Arnol'd}, in other words, as linear
superpositions of a few nearly degenerate true eigenstates. In many
situations, of course, a quasimode provides a correct and more physical
description of a phenomenon than the true eigenmodes.

\section{Acknowledgments}

Supported in part by the United States NSF grant DMR-9625549 and United
States-Israel Binational Science Foundation, grant 99800319. R.N. was
partially supported by the NSF grant DMR98-70681 and the University of
Kentucky. We thank Prof. Director Peter Fulde for hospitality at the
Max-Planck-Institut f\"ur Physik komplexer Systeme in Dresden, where some of
this work was done.

\end{document}